\documentclass[twocolumn,aps,prb,floatsfix,showpacs,superscriptaddress]{revtex4}
\usepackage{graphics}
\usepackage{amssymb,amsmath,amsfonts}
\usepackage{array}
\usepackage{dcolumn}
\usepackage{color}

\begin{document}

\title{Fast calculation of the electrostatic potential in ionic crystals by
  direct summation method}


\author{Alain Gell\'e} 
\author{Marie-Bernadette Lepetit}

\affiliation{CRISMAT, ENSICAEN-CNRS UMR6508, 6~bd. Mar\'echal Juin,
14050 Caen, FRANCE}

\date{\today}


\begin{abstract}
An efficient real space method is derived for the evaluation of the
Madelung's potential of ionic crystals. The proposed method is an
extension of the Evjen's method. It takes advantage of a general
analysis for the potential convergence in real space. Indeed, we show
that the series convergence is exponential as a function of the number
of annulled multipolar momenta in the unit cell. The method proposed
in this work reaches such an exponential convergence rate. Its
efficiency is comparable to the Ewald's method, however unlike the
latter, it uses only simple algebraic functions.
\end{abstract}

\maketitle


\section{introduction}

 Since 90 years a number of methods have been proposed to calculate
 the electrostatic potential in ionic crystals. These methods can be
 separated into two categories, the direct summation methods and the
 indirect summation ones.  The former uses a real space summation of
 the electrostatic potential generated by the ions within a finite
 volume ($\cal E$). However, when enlarging the volume $\cal E$, such
 partial summations are conditionally convergent. The convergence
 depends on the specific shape of $\cal E$.  In addition, when
 achieved, the convergence is quite slow.  The indirect summation
 methods do not present these drawbacks since the long range part of
 the potential is calculated in the reciprocal space.  Indeed, the
 summation is divided into two parts, a short range one, evaluated by a
 direct summation in real space and a long range one evaluated in
 the reciprocal space. Among these methods, the most widely used is
 the Ewald's method~\cite{Ewald}, which is actually considered as the
 reference for Madelung potential calculations.

 Despite its quality the Ewald's method is not easily usable in
 different domains of physics. This is for instance the case in
 clusters ab-initio calculations used for the treatment of strongly
 correlated systems, the study of diluted defects in materials or
 adsorbates or for QM/MM type of calculations. For this types of
 calculations, real space direct summation methods are used.  
 There is thus a need for efficient and accurate techniques
 for the determination of the Madelung potential in real space. 

 The convergence problems found in real space summation are linked to
 the shape of the summation volume, $\cal E$, and more specifically
 the charges at its surface. In order to insure the convergence
 of the summation, the surface charges are renormalized. Several
 methods have been proposed for this purpose. 

 The most common and simple one is the Evjen's
 method~\cite{Evjen}. This method uses a volume  $\cal E$, built from
 a finite number of crystal unit cells, and renormalizes the surface
 charges by a factor 1/2, 1/4 or 1/8 according whether the charge
 belong to a face, edge or corner of $\cal E$. This method insures, in
 most cases, the convergence of the electrostatic potential when $\cal
 E$ increases. However, in some cases such as the famous $\rm Cs Cl$ it
 does not converge to the proper value.~\cite{Evjen}

 Other authors~\cite{ajust_ch1,ajust_ch2} proposed to renormalize not only the
 surface charges, but also the charges included in a thin skin
 volume. The adjustment of the renormalization factors are, in this
 case, numerically determined so that to reproduce the exact
 potential, previously computed using the Ewald's method at a chosen
 set of positions. Such a method presents the advantage of reaching a
 very good precision. However several drawbacks can be pointed
 out~: i) the previous calculation of the electrostatic potential
 using the Ewald's method at a large number of space positions, ii)
 the necessity to invert a large linear system to determine the
 renormalization factors and iii) finally the fact that the latter are
 not chosen on physical criteria. Indeed, this last point induces the
 possibility that the renormalization factors can be either larger
 than one or negative. It results that even if the electrostatic
 potential is very accurate at the chosen reference positions, its
 spatial variations can be unphysical and thus, the precision can
 strongly vary when leaving the reference points.

 Marathe {\it et al}~\cite{Marathe} suggested a physical criterion,
 based on the analysis of the convergence of the real space summation,
 for the choice of the renormalization factors. Indeed, it is known
 that the direct space summation converges to the proper limit if the
 volume $\cal E$ presents null dipole and quadrupole
 momenta~\cite{Dahl,Coogan}. The authors of
 reference~\onlinecite{Marathe} showed, on the simple example of a
 linear alternated chain, that it is possible to find a finite number
 of charge renormalization factors allowing the cancellation of these
 two multipolar momenta. They also assert that the cancellation of
 additional multipolar momenta increases the speed of
 convergence. Unfortunately they did not prove this affirmation and
 more importantly, they did not proposed a practical way to determine
 the renormalization factors in order to reach this goal.
 
 In the present paper we propose a systematic method for the
 determination of the renormalization factors allowing the
 cancellation of a given number of multipolar momenta as well as a
 careful analysis of the direct space summation convergence as a
 function of the number of canceled multipolar momenta.  The next
 section will present the convergence proof, section 3 will develop
 the method for determination of the renormalization factors and
 section 4 will present the optimization of the method and 
 illustration on a typical example.

\section{Convergence analysis}
\label{s:Conv}
\subsection{Potential at a point}

As already mentioned in the introduction, several papers already exist on this
subject. However the results are only partial and there is not complete
analysis of the convergence issue. We will thus present in this section a
global analysis of the electrostatic potential convergence in a real space
approach and an estimation of the error.

We want to evaluate the limit of the following series
\begin{equation} 
V(\vec r_0) = \lim_{n\rightarrow \infty} V_n(\vec r_0) = 
\lim_{n\rightarrow \infty} \sum_{\vec R\in {\cal E}_n} \sum_{j\in {\cal C}} 
\frac{q_j}{|\vec R + \vec r_j - \vec r_0|}  
\end{equation}
where $\vec R$ is a vector of the Bravais's lattice, j refers to a charge
$q_j$ located at the position $\vec r_j$ of the unit cell $\cal C$. $\left\{
  {\cal E}_n, n\in \mathbb N\right\}$ is a set of volumes such that
\begin{equation}
 \lim_{n\rightarrow \infty} {\cal E}_n = \mathbb{R}^3
\qquad {\rm and} \qquad \vec r_0 \in {\cal E}_0
\end{equation}
For the sake of simplicity
we require that the set of ${\cal E}_n$ also presents the following conditions
\begin{eqnarray}
&&{\cal E}_0 \subset {\cal E}_1 \subset {\cal E}_2 \subset 
\dots \subset {\cal E}_n \\
 \text{and} && \nonumber \\
&& \text{if } \vec R \in {\cal E}_n \text{ then }  -\vec R \in {\cal E}_n
\end{eqnarray}

The well known problem of this series is that the limit depends on the
particular choice of the unit cell ${\cal C}$ and of the volumes
${\cal E}_n$.  However, it has been demonstrated that this conditional
convergence disappears if one considers a unit cell with zero dipolar
and quadrupolar momenta~\cite{Coogan}. 
We will consider in the following that the cell $\cal C$ fulfills this
condition.

The error $\Delta V_n(\vec r_0)$ on the electrostatic potential 
evaluation, $V_n(\vec r_0)$, can be written as 
\begin{equation}
  \Delta V_n(\vec r_0)=\sum_{\vec R\,\not\in\,\mathcal{E}_n}
  \sum_{j\,\in\,\mathcal{C}} \frac{q_j}{|\vec R+\vec r_{j}-\vec r_0|}
\end{equation}
For large values of $n$, $\Delta V_n(\vec r_0)$ can be evaluated by a
multipolar expansion. For practical reasons, we will use an expansion
expressed in spherical coordinates.  Indeed, for a given order, this
expansion contains less terms than the usual multipolar expansion
based on Cartesian coordinates. The use of the later multipolar
expansion is still possible but is more complex (see
ref.\onlinecite{mathese}). In spherical coordinates, the multipolar
expansion of the error made on $V_n\left(\vec r_0\right)$
reads~\cite{multiYlm}~:
\begin{equation}
  \Delta V_n\left(\vec r_0\right) = \sum_{\vec R\,\not\in\,\mathcal{E}_n}
  \sum_{l} \sum_{m=-l}^{l}  \mathcal{M}_{lm}(\vec r_0)
  \frac{Y_l^m(\theta,\phi)}{R^{l+1}} 
\label{eq:DVY}
\end{equation}
where $\mathcal{M}_{lm}(\vec r_0)$ are the multipolar momenta of unit
cell $\mathcal{C}$ at $\vec r_0$. They can be expressed as
\begin{equation}
  \mathcal{M}_{lm}(\vec r_0) = \sum_{j \in {\cal C}} q_j |\vec r_j - \vec
  r_0|^l Y_l^{-m}(\theta_j,\phi_j) 
\label{eq:multip}
\end{equation}
$(R,\theta,\phi)$ are the spherical coordinates of the Bravais's vector $\vec
R$ and $(\rho_j,\theta_j,\phi_j)$ the spherical coordinates of $\vec r_j - \vec
r_0$. The $Y_l^{-m}$ are Schmidt semi-normalized spherical harmonics~:
\begin{equation}
Y_l^{-m}(\theta_j,\phi_j)=\sqrt{\frac{(l-m)!}{(l+m)!}} P_l^m(\cos{\theta})e^{im\phi}
\end{equation}
where $P_l^m$ are Legendre functions.

In order to overvalue the error, one needs to  overvalue the spherical
harmonics. We thus consider the addition formula~:
\begin{equation}
  \sum_{m=-l}^{l} Y_l^m(\theta,\phi)\,Y_l^{-m}(\theta',\phi')=
  P_l(\cos\,\gamma) 
\end{equation}
where $P_l$ is a Legendre Polynomial and $\gamma$ is the angle between
$(R,\theta,\phi)$ and $(R',\theta',\phi')$. 
It comes for $\theta=\theta'$ and $\phi=\phi'$
\begin{equation}
  \sum_{m=-l}^{l} | Y_l^m(\theta,\phi) |^2 =  P_l(1) = 1
\end{equation}
and thus 
 \begin{equation}
   | Y_l^m(\theta,\phi) | < 1
\label{eq:majY}
\end{equation}

Using this result, one obtains the following overvalue for the momenta~:
\begin{equation} 
  |\mathcal{M}_{lm}| < Q  (r_0 +a)^l 
\label{eq:majM}
\end{equation}
where $a$ is the typical size of $\mathcal{C}$ (i.e. the diameter of
the circumsphere of $\mathcal{C}$) and $Q$ is the sum of the absolute
values of its charges
\begin{equation}
Q=\sum_{j \in \mathcal{C}} |q_j|
\end{equation}

One should notice at this point that $Y_l^m(\theta,\phi)$ has the same parity
as $l$.  The contributions of $\vec R$ and $-\vec R$ unit cells to $\Delta
V_n\left(\vec r_0\right)$ thus cancel when $l$ is odd. Using
equations~\ref{eq:majY} and~\ref{eq:majM}, one gets the following
overvaluation~:
\begin{equation}
  |\Delta V_n\left(\vec r_0\right)| < 
  Q\, \sum_{k=p}^{\infty} (4k+1)\, (r_0 +a)^{2k}
\sum_{\vec R\,\not\in\,\mathcal{E}_n}
\frac{1}{R^{2k+1}}  
\end{equation}
where $2p$ is the order of the first even, non-zero momentum.

Let us now overvalue the sum over the $1/R$ powers by a
volume integral.  For each cell ${\cal C}(\vec R)$ located at $\vec
R$, the norm of the position vectors $\vec r$ belonging to the volume
${\cal C}(\vec R)$ is smaller than $R+a$. One can thus overvalue
$1/R^{2k+1}$ by
\begin{equation}
  \iiint_{{\cal C}(\vec R)}
  \frac{1}{(r-a)^{2k+1}}\frac{dv}{\omega}
\end{equation}
$\omega$ being the volume of the unit cell $\mathcal{C}$. If ${\cal
R}_n$ is the radius of the insphere of ${\cal E}_n$, it comes
\begin{eqnarray}
\sum_{\vec R\,\not\in\,\mathcal{E}_n}\frac{1}{R^{2k+1}} &<& 
\iiint_{r>{\cal R}_n} \frac{1}{(r-a)^{2k+1}} \frac{dv}{\omega} \nonumber \\ 
&<& \frac{4\pi}{\omega} \, \frac{1}{2k-2} \,
\frac{{\cal R}_n^2 }{({\cal R}_n -a)^{2k}}
\label{eq:sumk_maj}
\end{eqnarray}
and
\begin{equation}
  |\Delta V_n\left(\vec r_0\right)| < \frac{4\pi Q}{\omega}\, 
  \sum_{k=p}^{\infty} 
  \frac{4k+1}{2k-2} \, {\cal R}_n^2 \, 
  \left( \frac{r_0 +a}{{\cal R}_n -a} \right)^{2k} 
\end{equation}

The later sum converges for ${\cal R}_n$ large enough, i.e. for ${\cal
  R}_n>r_0 +2a$.
The decreasing function $(4k+1)/(2k-2)$ can be overvalued by its value
in $k=p$. Further summation over $k$ leads to the following expression
\begin{equation} 
  |\Delta V_n\left(\vec r_0\right)| < 
  a_p\,  \frac{{\cal R}_n^2 (r_0 +a)^{2p}}{({\cal R}_n-r_0-2a)({\cal R}_n -a)^{2p-1}}
\label{eq:res1}
\end{equation}
with
\begin{equation}
a_p=\frac{4\pi\,Q}{\omega} \times \frac{4p+1}{2p-2} \,
\end{equation}
The electrostatic potential at $\vec r_0$ thus converges as $1/{\cal
R}_n^{l-2}$ where $l$ is the first, even, non-zero momentum of the
unit cell ${\cal C}$.

\subsection{Difference of potential between two points}

In several applications, as for instance in cluster ab initio
calculation, the problem depends on the spatial variations of the
potential and not on its absolute value. In such cases it is
sufficient to cancel the dipolar momentum of the unit cell in order to
ensure the convergence of the calculation. The convergence rate can
also be expected to be faster than for the calculation of the
potential at a point as we will show in this section. 

Let us overvalue the error made on the calculation of a difference of
potential between two points located at $\vec r_o+\vec r_1$ and
$\vec r_o-\vec r_1$~:
\begin{widetext}
\begin{eqnarray}
  \Delta V_n(\vec r_0,\vec r_1)&=&\sum_{\vec R\,\not\in\,\mathcal{E}_n}
  \sum_{j\,\in\,\mathcal{C}} \frac{q_j}{|\vec R+\vec r_{j}-\vec r_0-\vec r_1|}
  - \frac{q_j}{|\vec R+\vec r_{j}-\vec r_0+\vec r_1|} \nonumber\\
 &=& \sum_{\vec R\,\not\in\,\mathcal{E}_n}
  \sum_{L} \sum_{M=-L}^{L}  \mathcal{M}_{LM}(\vec r_0)
  \left( \frac{Y_L^M(\theta_-,\phi_-)}{ R_-^{l+1}}-
    \frac{Y_L^M(\theta_+,\phi_+)}{ R_+^{l+1}}\right)
\label{eq:dv2}
\end{eqnarray}
\end{widetext}
where $(R_-,\theta_-,\phi_-)$ and $(R_+,\theta_+,\phi_+)$ are the
spherical coordinates of $\vec R-\vec r_1$ and $\vec R+\vec r_1$
respectively. As in the preceding section $(R,\theta,\phi)$ will be
the spherical coordinates of $\vec R$ and $(r_1,\theta_1,\phi_1)$
those of $\vec r_1$.

In order to express the previous expression as a function of $1/R$, we use
following expansion of solid spherical harmonics (for simple derivation see
ref.~\onlinecite{expylm_Dahl}, see also
ref.~\onlinecite{expylm_Sack,expylm_Chiu})~:
\begin{eqnarray}
\frac{P_L^M\!(\cos{\theta_-})\, e^{iM\phi_\pm}}{ R_-^{L+1}}
&=& \sum_{n=0}^\infty \frac{r_1^n}{R^{L+n+1}} 
\sum_m \binom{L+n-m}{L-M}\nonumber\\*
&&\times\, (-1)^{M-m} P_{L+n}^m\!(\cos{\theta})\, e^{im\phi}\nonumber\\*
&&\times\,   P_{n}^{M-m}\!(\cos{\theta_1})\, e^{i(M-m)\phi_1} 
\end{eqnarray}
where the sum over $m$ spans all integer values. Nevertheless, only a
finite number of terms will contribute, since $P_l^m=0$ if
$|m|>l$. Setting $l=L+n$ and introducing spherical harmonics leads
to~:
\begin{eqnarray}\label{eq:ylmexp}
  \frac{Y_L^M(\theta_-,\phi_-)}{ R_-^{L+1}} &=& 
  \sum_{l=L}^\infty 
  \sum_m  \frac{Y_{l}^{m}(\theta,\phi)}{R^{l+1}} \nonumber\\* 
  &&\times\, (-1)^{M-m} \left[\binom{l-m}{L-M}
    \binom{l+m}{L+M}\right]^{\frac{1}{2}}  \nonumber\\* 
  &&\times\, r_1^{l-L}\, Y_{l-L}^{M-m}(\theta_1,\phi_1)
\end{eqnarray}
Considering $\vec R_+$ in this equation instead of $\vec R_-$ is
equivalent to the transformation 
\begin{eqnarray*}
  \theta_1 &\longrightarrow &  \pi - \theta_1 \\
  \phi_1 &\longrightarrow & \pi + \phi_1
\end{eqnarray*}
that results in an overall $(-1)^{l-L}$ factor. Inserting
relation~\ref{eq:ylmexp} into eq.~\ref{eq:dv2} and inverting the summation
over $l$ and $L$ leads to the expansion~:
\begin{equation}
  \Delta V_n(\vec r_0,\vec r_1) = 
  \sum_{\vec  R\,\not\in\,\mathcal{E}_n}
  \sum_{l=0}^{\infty}\sum_{m=-l}^l
  \frac{Y_{l}^{m}(\theta,\phi)}{R^{l+1}} A_{lm}
\label{eq:dv2_alm}
\end{equation}
with~:
\begin{eqnarray}
  A_{lm}&=&\sum_{L=0}^l \sum_{M=-L}^L  \mathcal{M}_{LM}(\vec r_0)
  \left[\binom{l-m}{L-M} \binom{l+m}{L+M}\right]^{\frac{1}{2}}  \nonumber\\* 
  &&\times (-1)^{M-m} \; r_1^{l-L}\; Y_{l-L}^{M-m}(\theta_1,\phi_1) \nonumber\\* 
  &&\times \left(1-(-1)^{l-L} \right)
\label{eq:alm}
\end{eqnarray}

Considering the parity of $Y_{lm}$, one can see from
eq.~\ref{eq:dv2_alm} that the contributions from cells located at
$\vec R$ and $-\vec R$ cancel when $l$ is odd.  Moreover, due to the
last term of eq.~\ref{eq:alm}, the $A_{lm}$ coefficients are zero when
~$l-L$~ is even. Only terms with $l$ even and $L$ odd have a non zero
contribution, thus only momenta $\mathcal{M}_{LM}(\vec r_0)$ with odd
order will contribute to the error. The consequence is that the first
non-zero contribution in equation~\ref{eq:dv2_alm} corresponds to 
$l=2p+2$ where $2p+1$ is the first, non-zero, odd momentum of the unit
cell.

Let us now find an overvalue of the $A_{lm}$ terms. It is easy to show
using a recurrence relation on the values of $m$ and $M$, that if
$|m|\le l$, $|M|\le L$ and $|M-m|\le l-L$, the following relation
holds~:
\begin{equation}
\binom{l-m}{L-M}\binom{l+m}{L+M} \le \binom{l}{L}^2
\end{equation}
Using the previous overvaluation of the momenta (eq.~\ref{eq:majM}) one
obtains~:
\begin{eqnarray}
 | A_{lm}|&\leq& 2Q\sum_{L=0}^l (2L+1)\binom{l}{L}(r_0+a)^L r_1^{l-L}
  \nonumber\\*
&\leq& 2Q(2l+1)(r_0+a+r_1)^l
\end{eqnarray}

As in previous section, the sum over ${\cal R}$ can be overvalued by a volume
integral (cf. eq~\ref{eq:sumk_maj}). Overvaluation of the error thus reads~:
\begin{equation}
|\Delta V_n(\vec r_0,\vec r_1)| \leq \sum_{k=p+1}^\infty\!
\frac{8\pi Q(4k+1)^2\, {\cal R}_n^2}{\omega(2k-2)}
  \left( \frac{r_0+a+r_1}{{\cal R}_n -a}\right)^{2k}
\end{equation}

The sum over $k$ converges if ${\cal R}_n$ is larger than $r_0 +2a+r_1$. It can
be calculated using derivative of power series. After simplification, one
obtains~:
\begin{equation}
 |\Delta V_n(\vec r_0,\vec r_1)| \leq b_p    \frac{{\cal
      R}_n^2 (r_0 +a+r_1)^{2p+2}}{({\cal R}_n-r_0-2a-r_1)^2({\cal R}_n
    -a)^{2p}} 
\label{eq:res2}
\end{equation}
where
\begin{equation}
b_p=\frac{8\pi Q}{\omega}\times\frac{16p^2+40p+25}{2p}
\end{equation}
As one increase the size of the set of charges, the difference of
electrostatic potential between two points converges like $1/{\cal
  R}_n^{l-1}$, where $l$ is now the first, odd, non-zero momentum of the unit
cell ${\cal C}$. This convergence is slightly faster than the convergence of
the absolute value of potential. When the order of the first non zero momentum
is even, the convergence rates differ by a factor $1/R^2$, otherwise they are
similar.


\section{Partial charges}

As depicted in the previous section, convergence can be considerably
increased if one cancels several multipolar momenta of the unit
cell. In general the Evjen method allows to only cancel the dipolar
momentum, and thus provides a convergence of the potential differences
in $1/{\cal R}_n^{2}$. In order to really take advantage of the former
property, one needs a method allowing the cancellation of several
multipolar momenta.  

In this section we will establish a method to construct unit cells
with a chosen number of zero multipolar momenta. The method, based on
the usage of partial charges, is general and can be applied to any
Bravais's crystal.

Let $(\vec a_0,\vec b_0,\vec c_0)$ be the lattice vectors of the
Bravais's crystal, and ${\cal C}_0$ the associated unit cell. In order
to introduce partial charges, we consider a larger cell ${\cal C}_l$
of dimensions $(l\times a_0,l\times b_0,l\times c_0)$, that we will
refer as the ``{\em construction cell}''. The construction cell thus 
contains $l^3$ original unit cells ${\cal C}_0$ which positions in
${\cal C}_l$ can be labeled by $p$, $q$ and $r$ indices, ranging from
$1$ to $l$.


If we note $n_c$ the number of charges $q_i$ in the original cell
  ${\cal C}_0$, the cell ${\cal C}_l$ now contains $n_c\times l^3$
  charges. These charges will be corrected by a factor
  $\lambda_{pqr}^i$ (where $i$ refers to the charge $q_i$). When on
  rebuild the lattice using the construction cells ${\cal C}_l$, the
  cells overlap, and the final charge at position $\vec r_i$ 
  corresponds to the superposition of partial charges from several
  construction cells. It is straightforward to show that the condition
  to retrieve the nominal value of the charges $q_i$ reads ~:
\begin{equation}
 \sum_{p,q,r=1}^l \lambda_{pqr}^i = 1 \qquad (1\leq i\leq n_c  
\label{eq:recon1}
\end{equation}

At this stage, considering the latter $n_c$ equations, the cell ${\cal
C}_l$ contains $n_c(l^3-1)$ free parameters that could be used to
cancel multipolar momenta. For the sake of simplicity and generality
(i.e. for the method not to depend on the particularity of a given
crystal), we will impose further conditions on the $\lambda_{pqr}^i$
coefficients.

We first reduce the problem to a one dimensional problem by setting~:
\begin{equation}
\lambda_{pqr}^i =\lambda_{p}^{a,i}\,\lambda_{q}^{b,i}\,\lambda_{r}^{c,i}
\end{equation}
where the three coefficients $\lambda_{p}^{a,i}$, $\lambda_{q}^{b,i}$
and $\lambda_{r}^{c,i}$ are used to cancel multipolar momenta of the
$1D$ problems obtained when the cell ${\cal C}_0$ is respectively
projected on the three axes of the crystal. For each one dimensional
problem, the construction cell contains $l$ projected unit cells. The
condition on the coefficients now reads~:
\begin{equation}
\sum_{p=1}^l \lambda_{p}^{\omega,i} = 1\qquad (\omega=a,b,c)
\label{eq:recon2}
\end{equation}
It is easy to show that, if these coefficients cancel a fixed number of
multipolar momenta in each one dimensional problems, the $\lambda_{pqr}^i$
coefficients will cancel the momenta of same order in the original three
dimensional problem.

\begin{figure*}[hbtp] 
\resizebox{12cm}{!}{\includegraphics{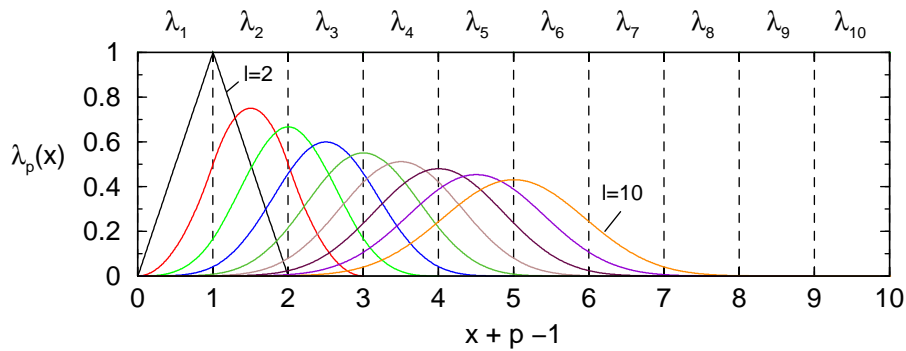}}
\caption{(color online) $\lambda_p(x)$ functions obtained for values of $l$
  ranging between $2$ and $10$. Intervals correspond to cells ${\cal C}_0$
  that compose the construction cell ${\cal C}_l$. In each intervals,
  fractional coordinates $x$ range between $0$ and $1$.}
\label{f:lp}
\end{figure*}

We further impose to the coefficients to only depend on the fractional
coordinates ($\alpha_i,\beta_i,\gamma_i$) of the charges $q_i$, in the
corresponding direction:
\begin{equation}
  \lambda_{pqr}^i
  =\lambda_{p}(\alpha_i)\,\lambda_{q}(\beta_i)\,\lambda_{r}(\gamma_i) 
\label{eq:l3}
\end{equation}
The $\lambda_{p}(x)$ functions are thus the same for the three directions and
for all charges. As a consequence their expression is the same for all
crystals.
For a given value of $x$, and considering the condition for the
reconstruction of the crystal (eq.~\ref{eq:recon2}), we are left with
$l-1$ degrees of freedom.  We thus impose to the $\lambda_{p}(x)$
functions to cancel $l-1$ multipolar momenta. This can be done by
setting the momenta created at the center of the construction cell by a unique 
charge $q$~:
\begin{eqnarray} \label{eq:mk}
q \; u_0^k \; \sum_{p=1}^l 
\lambda_{p}(x)\left(x+p-1-\frac{l}{2}\right)^k&=&q \; u_0^k \; m_{l,k} \\
&& (0\leq k\leq l-1) \nonumber
\end{eqnarray}
where $u_0=a_0,b_0,\text{ or }c_0$, $x$ is the fractional
coordinate of the charge and $m_{l,k}$ are constant values
(independant of the crystal specifications). The equation obtained for $k=0$
corresponds to the condition for the reconstruction of the crystal
($m_{l,0}\equiv 1$).  The momenta of the construction cell, can thus
be obtained by summing the contributions of all charges. As the unit
cell is neutral, these contributions cancel out.

The equations~\ref{eq:l3} and~\ref{eq:mk} thus define sets of partial
charges that allow to construct cells with $l-1$ zero multipolar
momenta.  The shape of these partial charges depends on the choice of
the $m_{l,k}$ constants values. In order to find the most reasonable
choice of partial charges, we search for $m_{l,k}$ constants that
satisfy the following physical conditions~:
\begin{enumerate}
\item $\forall p, \; \lambda_p(x) \in [0,1]$.
\item The partial charges vary continuously, i.e. $\forall p, \;
  \lambda_p(x)$ are continuous and $\lambda_p(1)~=~\lambda_{p+1}(0)$.
\item $\forall p,\; \lambda_p(x)$ decreases monotonously when moving
  away from the center of the construction cell.
\item The values of the $m_{l,k}$ are as small as possible.
\end{enumerate}
The latter condition ensures that the partial charges are larger in the center
of the construction cell and smaller on its edges, and hence that this cell is
close to the original one.

We did not find a way to derive the solution of this problem in a
general way, for any value of $l$.  We thus determined the solution
for fixed values of $l$ up to $l=6$. In all these cases the first
$\lambda_p$ function presents the same shape~:
\begin{equation}
  \lambda_1(x)=\frac{x^{l-1}}{(l-1)!}
\label{eq:l1}
\end{equation}
We reasonably assume that this expression is valid for any value of $l$. As we
will see, it is possible to show, a posteriori, that the $\lambda_p$ functions
fulfill the first three conditions. 

We will now determine the function $\lambda_p$ for any $l$, using the
above expression of $\lambda_1$ and relation~\ref{eq:mk}. In
equation~\ref{eq:mk}, the $m_{k,i}$ constants can be replaced by the
value of the momenta obtained for $x=0$~:
\begin{equation}
  \sum_{p=1}^l \lambda_{p}(x)\left(x+p-1-\frac{l}{2}\right)^k=
  \sum_{p=1}^l \lambda_{p}(0)\left(p-1-\frac{l}{2}\right)^k
\label{eq:mk2}
\end{equation}
with $0\leq k\leq l-1$. The matrix of this linear system is the transpose of a
Vandermonde Matrix. Inversion of the system leads to~:
\begin{equation}
 \lambda_{p}(x) = \sum_{q=1}^l \lambda_{q}(0)\prod_{i=1,i\neq p}^l \frac{x+i-q}{i-p}
\label{eq:lpl0}
\end{equation}

The $\lambda_p(0)$ coefficients can now be obtained from the expression of
$\lambda_1$. Let us consider the latter equation evaluated for $p=1$, and for
$l$ integer values of $x$~:
\begin{equation}
\lambda_1(j) = \sum_{q=1}^l \lambda_{q}(0) \binom{l+j-q}{j-q+1} \qquad (1\leq
j \leq l)
\end{equation}
Replacing $\lambda_1(n)$ by its expression and inverting this linear system
leads to the expression of $\lambda_q(0)$ coefficients~:
\begin{equation}
 \lambda_{q}(0)=\sum_{j=1}^{q-1} (-1)^j\binom{l}{j} 
\frac{(q-j-1)^{l-1}}{(l-1)!}
\label{eq:l0}
\end{equation}

\begin{figure*}[hbtp] 
\resizebox{12cm}{!}{\includegraphics{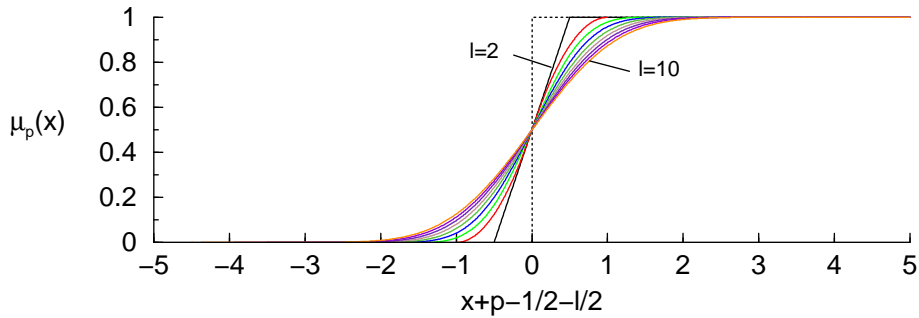}}
\caption{(color online) Renormalization of the charges at the ``left'' edge of
  a crystal fragment, obtained for values of $l$ ranging between $2$
  and $10$. The origin of the abscises corresponds to the position of
  the edge obtained when using the original cell ${\cal C}_0$ (dotted line).}
\label{f:mp}
\end{figure*}

Finally, using eq.~\ref{eq:lpl0} and~\ref{eq:l0} one obtains the expression of
the $\lambda_p(x)$ which are polynomial functions of order $l-1$ in $x$ (see
fig.~\ref{f:lp})~:
\begin{equation}
  \lambda_{p}(x) = \sum_{q=1}^l\sum_{j=1}^{q-1} (-1)^j \binom{l}{j}
  \frac{(q-j-1)^{l-1}}{(l-1)!} \!\prod_{i=1,i\neq p}^l\! \frac{x+i-q}{i-p} 
\end{equation}
These functions are segments of the uniform sum distribution , i.e. the
distribution $P_l(x)$ of the sum of $l$ uniform variates on the interval
$[0,1]$~:
\begin{equation}
\lambda_p(x)=P_l(x+p-1)\qquad (0\leq x\leq 1 {\rm \ and\ } 1\leq p\leq l-1)
\end{equation}
From this relation, it is obvious that the $\lambda_p(x)$ functions
satisfy the first three conditions mentioned above.

We now consider a fragment of crystal made of $n$ construction cells ${\cal
  C}_l$. As these cells are composed of $l$ original cells ${\cal C}_0$, they
partially overlap, and the size of the fragment corresponds to $n+l-1$ cells
${\cal C}_0$.  $n'=n-l+1$ cells ${\cal C}_0$ in the center contains charges
with the nominal values $q_i$, and $l-1$ cells ${\cal C}_0$ on each side of the
fragment contains partial charges.  The latter partial charges are
proportional to the coefficients~:
\begin{equation}
\mu_p(x)=\sum_{q=1}^p \lambda_q(x)\qquad (1\leq p\leq l-1)
\end{equation}
These coefficients are represented on fig.~\ref{f:mp}. The abscise
values have been shifted by $(l-1)/2$, so that the origin corresponds
to the position of the effective edge of the fragment (i.e. the
position of the edge obtained when using $n$ original cells ${\cal
C}_0$ without partial charges).  One can see that the renormalization
of the charges is relatively small. Indeed, even in the case $l=10$,
this renormalization is weaker then $1\%$ for charges at  distances
larger than $2u_0,\; (u_0=a_0,b_0\text{ or }c_0)$.

\section{Optimized method}

\begin{figure}[hbtp] 
\resizebox{!}{5.5cm}{\includegraphics{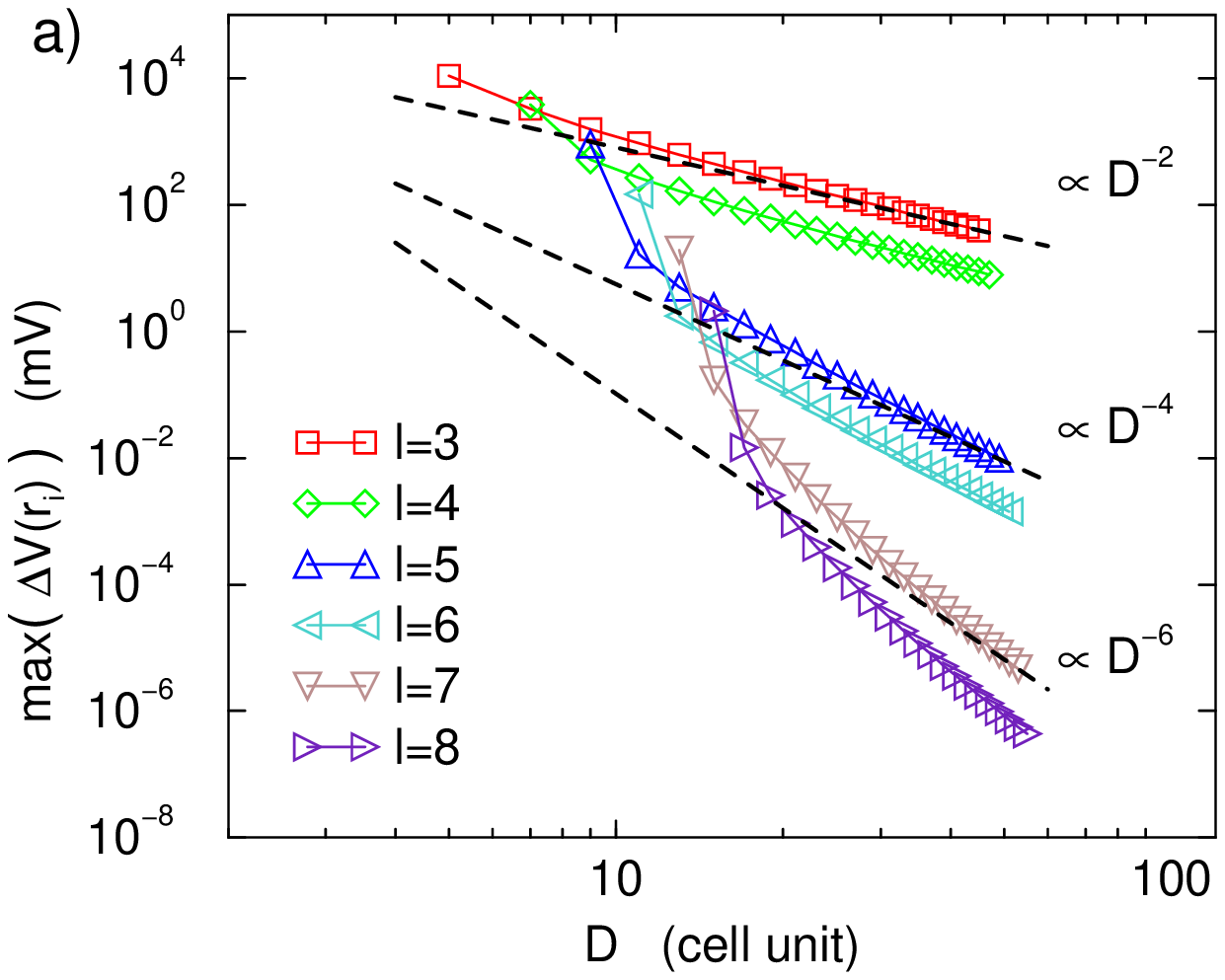}}\\
\resizebox{!}{5.5cm}{\includegraphics{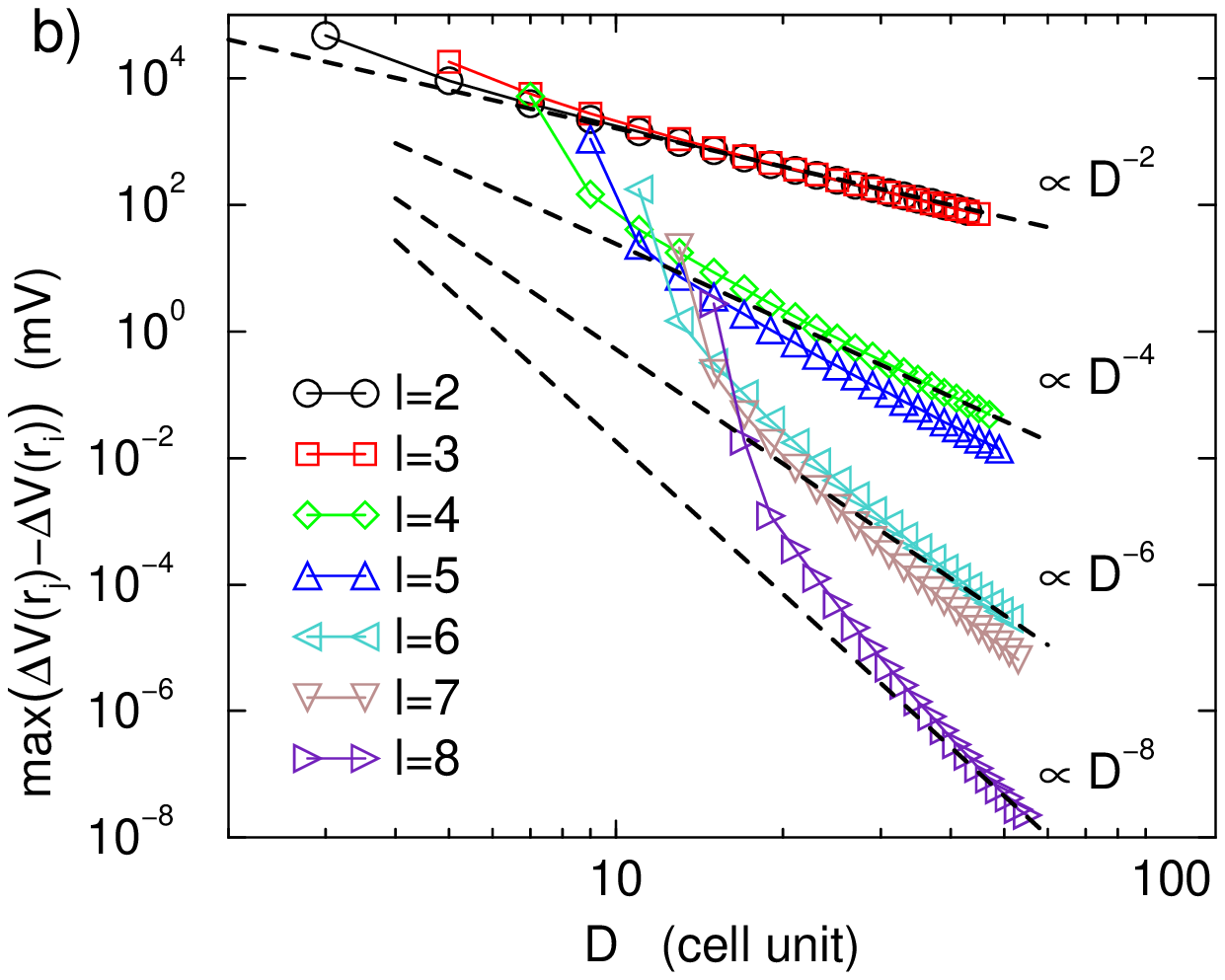}}
\caption{(color online) a) Maximum of error made on the calculation of
  the potential at the position of the atoms of the central unit
  cell. $D=n'+2l-2$ is the width of the sets of charges in each
  crystallographic direction, where $n'$ is the width of the central
  zone where the charges have their nominal values. $l$ is the order
  of the first non zero momenta of the cell ${\cal C}_l$ used to
  construct the set of charges. b) Maximum of error made on the
  calculation of the difference of potential.  }
\label{f:dv}
\end{figure}

We first use the cells ${\cal C}_l$ to illustrate the convergence of
the standard, real-space calculation of the potential established in
section~\ref{s:Conv}. In order to observe the general behavior of the
different methods, we chose the $\alpha$ quartz structure, that
possesses a reasonable number of atoms per unit cell and not too many
symmetries. A cell ${\cal C}_l$ is constructed for a fixed number of
$p$. The cells are used to produce sets of charges of increasing
size. The potential is calculated at the position of the twelve atoms
of the central unit cell ${\cal C}_0$.

Fig.~\ref{f:dv} a) represents the maximum of the error made on these
potential values as a function of the width of the set of charges, and
fig.~\ref{f:dv} b) represents the maximum of the error made on the
sixty six differences of potential. As expected the error decreases as
a power function of the width $D$ of the set of charges. It fully
agrees with eq.~\ref{eq:res1} and~\ref{eq:res2}. In particular, the
fact that the cancellation of a momentum of odd order do not increase
the convergence rate of the potential at a point, clearly
appears. Similarly, the cancellation of even order momentum do not
improve the convergence rate of the calculation of differences of
potential.

\begin{figure}[hbtp] 
\resizebox{!}{5.5cm}{\includegraphics{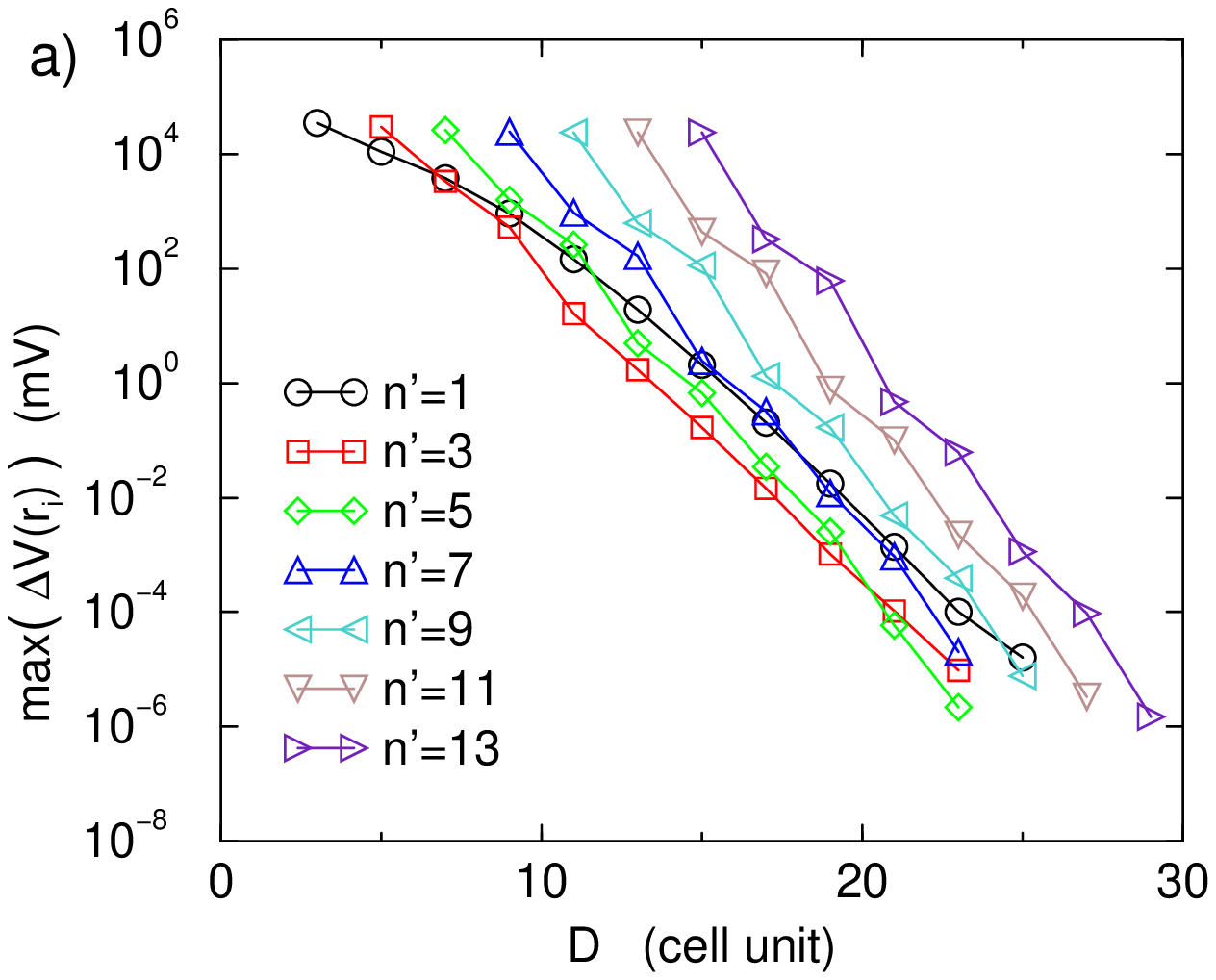}}\\
\resizebox{!}{5.5cm}{\includegraphics{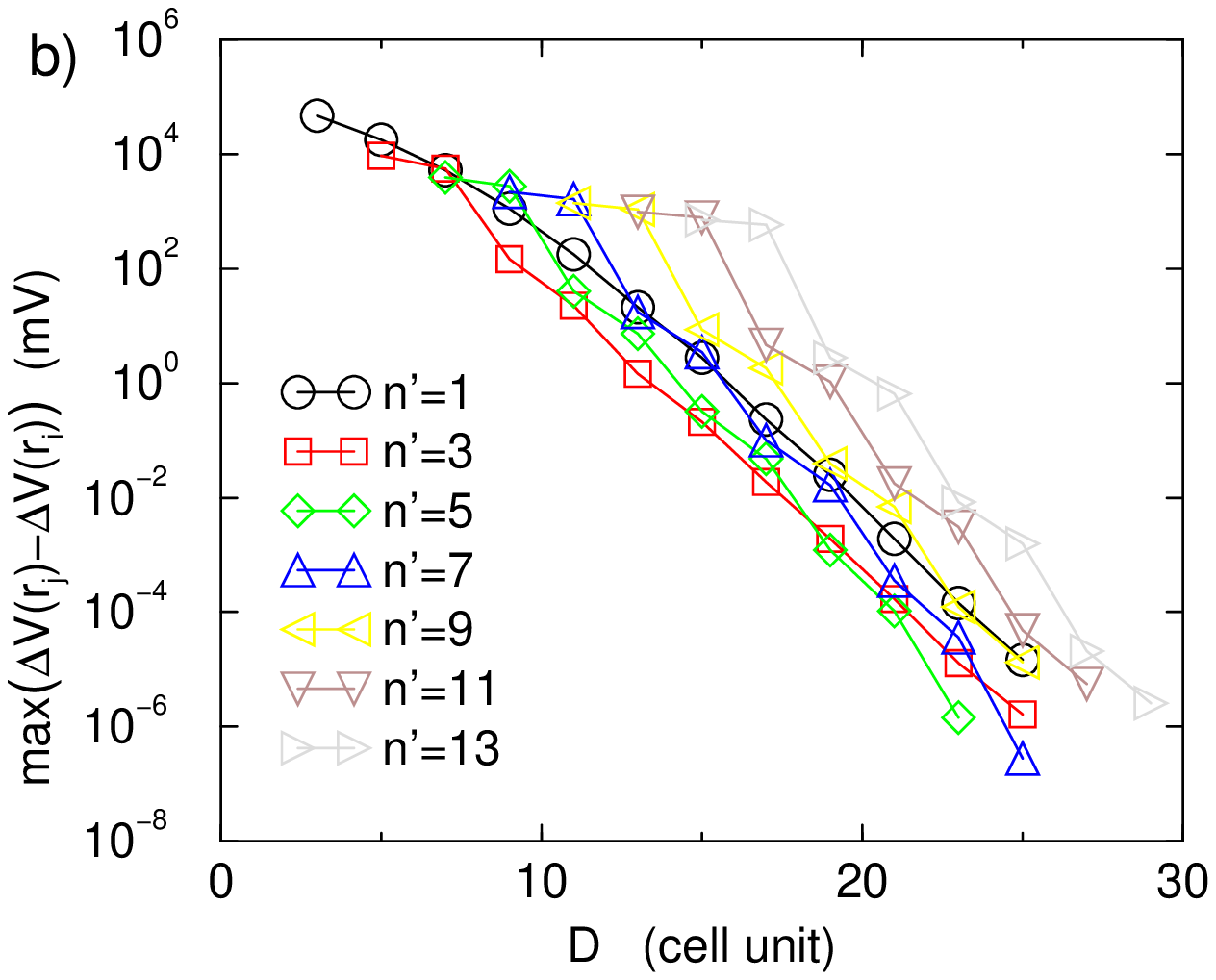}}
\caption{(color online) a) Maximum of error made on the calculation of the
  potential at the position of the atoms of the central unit cell. The abscise
  corresponds to the width $D=n'+2l-2$ of the sets of charge in each
  crystallographic direction, where $n'$ is the width of the central zone with
  nominal charge values. b) Maximum of error made on the calculation of the
  difference of potential.}
\label{f:dvexp}
\end{figure}

One can see from the previous figures that this standard approach is
not the more efficient. Indeed the increase of the number of zero
multipolar momenta clearly yield a faster convergence rate than the
increase of the volume of the system for a fixed value of $l$.  Let us
therefore fix the width $n'$ of the volume containing the nominal
charge, and let $l$ increase. The variation of the maximum error made
on the potential and on the the potential differences are respectively
represented on fig.~\ref{f:dvexp} a) and~\ref{f:dvexp} b). One sees
that the present approach leads to an exponential convergence of the
potential in both cases. The convergence is very fast, since an
increase of the set of charges width by two unit cells results in a
precision increase by a factor better than  $\sim 10$. Increasing the
number $n'$ of central cells without partial charges has a small
influence on the convergence speed.  For $n'>3$ the method even
becomes less efficient since increasing $n'$ increases the size of the
total set of charges. The best convergence is obtained for $n'=3$. It
corresponds to the case where the cell in which the potential is
calculated is surrounded by one shell of cells with the nominal charge
values.

\begin{figure}[hbtp] 
\resizebox{8cm}{!}{\includegraphics{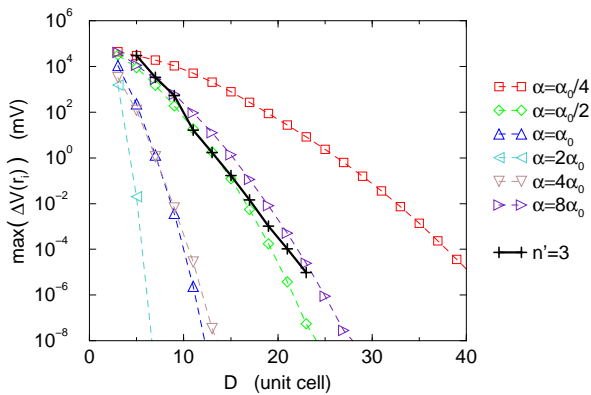}}
\caption{(color online) Error on the Madelung's potential evaluation
using the present method (bold black solid curve) and the Ewald's
method with different parameters $\alpha$ (colored dashed curves).}
\label{f:dvewald}
\end{figure}

Finally we compare our method to the famous Ewald's method which mixes
calculation in real space and reciprocal space. This method introduces
Gaussian distribution of charge $exp(-\alpha^2 r^2)$, where the $\alpha$
coefficient can be adjusted. Increasing $\alpha$ coefficient
increases the convergence rate of the real space sum, but slows down
the sum in reciprocal space. A width of Gaussian proportional to the
characteristic length of the unit cell, which corresponds to
$\alpha_0=\omega^{-1/3}$, is generally assumed to give a good
compromise. We calculated the error made on the value of the
potential using the Ewald's method for different values of $\alpha$ 
around $\alpha_0$.  The results are represented on
figure~\ref{f:dvewald}, as well as the error of our method obtained
for $n'=3$.
Figure~\ref{f:dvewald} reports the error on the potential as a
function of the number of construction cells used in the
calculation. Let us point out that, while this variable is pertinent
for the global convergence rate analysis, for each charge, the Ewald's
method requires an error function evaluation resulting in a non
negligible pre-factor, not present in our method and not taken into
account in figure~\ref{f:dvewald}. One sees that the convergence rate
of the present method is comparable with the Ewald's method.  If one
is only interested in the potential evaluation at a single point, the
Ewald's method with an optimal $\alpha$ parameter is somewhat faster
than the present one. One the other hand, once the renormalization
have been computed, the value of the potential at any other point of
the of the central area can be calculated with a similar precision at
little cost. More important, properties using potential integrals or
complex potential functions can be more easily evaluated since our
method used only algebraic functions.

\section{Conclusion}
Number of authors have searched for a fast converging method for the
evaluation of the electrostatic potential in real space. Similarly,
many works where done yielding partial results on the convergence rate
of such real series. The present work fills the gaps and proposes a
general analysis of both the convergence of the potential at one point
and of the convergence of differences of potential. Indeed, we gave a
general and rigorous proof of the relation (claimed by other
authors) between the power law convergence of the series and the
number of zero multipolar momenta of the crystal construction cell.

Based on these convergence analyses we derived a general real space
method with an exponential convergence rate, comparable with the
Ewald's method. The exponential convergence is reached as a function
of the number of canceled multipolar momenta in the {\em
construction} cell. The crystal is indeed constructed using
overlapping {\em construction} cells with renormalized charges.  We
derived a general analytical expression of the renormalization
factors, for any given number of zero multipolar momenta.



%
\end{document}